\documentclass[twocolumn,prb,showpacs,amsmath,amssymb,english]{revtex4}
\usepackage{graphicx}
\usepackage{dcolumn}
\begin{document}

\title {Multiple electron-hole scattering effect on quasiparticle properties \\in a homogeneous electron gas}

\author{I. A. Nechaev$^{1,}$}
\altaffiliation[Also at: ]{Theoretical Physics Department, Kostroma State University, 156961 Kostroma,
Russia.}
\author{E. V. Chulkov$^{1,2}$}
\affiliation{$^1$Donostia International Physics Center (DIPC), P. de Manuel Lardizabal 4, 20018, San
Sebasti{\'a}n, Basque Country, Spain\\
$^2$Departamento de F{\'\i}sica de Materiales, Facultad de Ciencias Qu{\'\i}micas, UPV/EHU and Centro Mixto
CSIC-UPV/EHU, Apdo. 1072, 20080 San Sebasti\'an, Basque Country, Spain}

\date{\today}

\begin{abstract}
We present a detailed study of a contribution of the $T$ matrix accounting for multiple scattering between an
electron and a hole to the quasiparticle self-energy. This contribution is considered as an additional term
to the $GW$ self-energy. The study is based on a variational solution of the $T$-matrix integral equation
within a local approximation. A key quantity of such a solution, the local electron-hole interaction, is
obtained at the small four-momentum transfer limit. Performed by making use of this limit form, extensive
calculations of quasiparticle properties in the homogeneous electron gas over a broad range of electron
densities are reported. We carry out an analysis of how the $T$-matrix contribution affects the quasiparticle
damping rate, the quasiparticle energy, the renormalization constant, and the effective mass enhancement. We
find that in comparison with the $GW$ approximation the inclusion of the $T$ matrix leads to an essential
increase of the damping rate, a slight reduction of the $GW$ band narrowing, a decrease of the
renormalization constant at the Fermi wave vector, and some ``weighting'' of quasiparticles at the Fermi
surface.
\end{abstract}

\pacs{71.10.-w}

\maketitle

\section{Introduction}

It is well known that the extensively used $GW$ approximation (GWA) to the quasiparticle self-energy is one
of the most successful methods for describing the quasiparticle properties in a broad spectrum of
materials.\cite{GW_review,Campillo,Zhukov_linear,Faleev} This approximation employs only the first term in
the Hedin self-energy diagrammatic expansion\cite{Hedin} in the dynamically screened Coulomb interaction (or
the so-called \textit{test charge---test charge} screened interaction). However, in a number of cases, for
example, ferromagnetic transition metals,\cite{FerdiAnis} the GWA is less satisfactory. It requires the cases
to go beyond the GWA (see, e.g., Refs.~\onlinecite{Zein_Antropov,Biermann,Ferdi3}).

One of the approximations that allows one to relatively simply go beyond the GWA is the so-called $GW\Lambda$
approximation ($\Lambda$ stands for the three-point vertex function). The latter includes vertex corrections
to the $GW$ self-energy in the same way as it can be done for the irreducible polarizability by means of the
spin-symmetric local-field factor. Thus the $GW\Lambda$ approximation, as well as the GWA, accounts for
charge-density fluctuations only. Moreover, this approximation depends weakly on the local-field factor and
gives results close to those in the GWA (see Refs.~\onlinecite{MahanGWGamma} and \onlinecite{Gurtubay}). Note
that it can be considered as a $GW$-like approximation, where the $GW$ formula for the quasiparticle
self-energy is used but with the \textit{electron---test charge} screened interaction instead of the
\textit{test charge---test charge} one.

The $GW$ formula for the self-energy can also be derived in approximations using the exact Ward identity as a
starting point (see, e.g., Ref.~\onlinecite{Local_approx}). In this case, the role of the screened
interaction $W$ is played by the effective \textit{electron---electron} screened interaction that includes
the local-field effects by means of the spin-symmetric and spin-asymmetric local-field factors and, thereby,
takes into account the contributions of both charge and spin fluctuations (see also
Refs.~\onlinecite{LA_Kukkonen,LA_Kukkonen_like,Nagy,Asgari}, and references therein).

The specific feature of the approaches mentioned above is the use of the local-field factors which are
defined outside the scope of these approaches and, as a rule, are tabulated and parametrized by using quantum
Monte Carlo (QMC) calculations for the homogeneous electron gas. It causes certain difficulties in the
description of quasiparticle properties, since these factors do not include real system band structure
effects. Thus, it is important to find a feasible and all-sufficient scheme for approximate calculations of
the self-energy which would allow us to preserve the advantages of the GWA and at that to give a possibility
of an inclusion of both charge- and spin-density fluctuations, that is crucial\cite{Zhukov,Asgari,Schafer} to
obtain agreement with experimental data.

In the preceding paper,\cite{IAN_EVC} hereafter referred to as I, we have examined a possibility to go beyond
the GWA by summing an infinite number of ladder diagrams of the Hedin self-energy expansion. To this end, we
have found a variational solution of the Bethe-Salpeter equation (more precisely the ladder approximation to
this equation\cite{FettWal}) determining the four-point $T$ matrix that describes multiple scattering of
propagating particles. The solution has been obtained within a local approximation in the spirit of
Refs.~\onlinecite{Local_approx} and \onlinecite{Richardson}. Making use of this solution, we have proposed a
form for the quasiparticle self-energy which allows one to take into account charge- and spin-density
fluctuations without double counting. A key quantity of such an approach is the local interaction
$\widetilde{W}$. In fact, the sum of the ladder diagrams is reduced to this interaction just as vertex
corrections to the random phase approximation (RPA) polarizability can be reduced to the spin-symmetric
local-field factor (see, e.g.,
Refs.~\onlinecite{Richardson,Tsolakidis,LFF_Shturm,LFF_TokPan,Suehiro_Ousaka_Yasuhara}). However, the point
is that the interaction $\widetilde{W}$ is defined by an eightfold integral and as such is problematical to
be realized in \textit{ab initio} calculations for real systems.

Thus, the goal of this paper is twofold. Having restricted ourselves to the multiple electron-hole scattering
case, first, we find an expression for $\widetilde{W}$ suitable for \textit{ab initio} calculations. As we
are mainly interested in small quasiparticle excitation energy, we consider the small four-momentum transfer
limit for this interaction. Second, making use of the obtained limit form of $\widetilde{W}$, we analyze what
effect the $T$-matrix contribution being an additional term to the $GW$ self energy has on quasiparticle
properties determined by both the self-energy and its derivatives with respect to momentum and frequency. In
this paper, we examine the $T$-matrix contribution as applied to the homogeneous electron gas in the
paramagnetic state.

The paper is organized as follows. In Sec.~\ref{sec:ladder_approx}, we derive explicit expressions for the
spin-diagonal and spin-non-diagonal parts of the local electron-hole interaction at the small four-momentum
transfer limit. Within the model of the homogeneous electron gas in the paramagnetic state, by making use of
the connection between the local interaction and the exchange part of the local-field factor, we examine
these expressions by comparing with the results known from the literature. In Sec.~\ref{sec:qp_properties} we
present our main results of extensive calculations carried out for quasiparticle properties over a broad
range of electron densities (for $r_s$ values ranging from 2 to 56). On the base of these results, we analyze
how the $T$-matrix inclusion with the obtained limit form for the local interaction modifies quasiparticle
properties evaluated from the $GW$ calculations. Finally, the conclusions are given in
Sec.~\ref{sec:conclusions}. Unless stated otherwise, atomic units are used throughout, i.e., $e^2=\hbar=m=1$.

%++++++++++++++++++++++++++++++++++++++++++++++++++++++++++++++++++++++++++++++++++++++++++++++++++++++++
\section{\label{sec:ladder_approx}Ladder approximation}
%++++++++++++++++++++++++++++++++++++++++++++++++++++++++++++++++++++++++++++++++++++++++++++++++++++++++
In this section, we derive basic expressions defining the local electron-hole interaction at the small
four-momentum transfer limit. Using the relation between this interaction and the exchange part of the
local-field factor, we compare our results for the homogenous electron gas (HEG) with those existing in the
literature.
%++++++++++++++++++++++++++++++++++++++++++++++++++++++++++++++++++++++++++++++++++++++++++++++++++++++++
\subsection{\label{sec_theory:T_matrix}$T$ matrix}
\begin{figure}[tbp]
\centering
 \includegraphics[angle=0,scale=0.7]{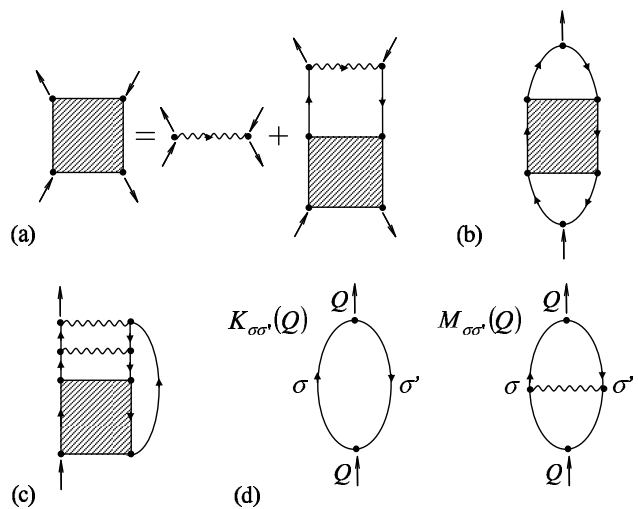}
\caption{A diagrammatic representation of the $T$ matrix (a), accounting for multiple electron-hole
scattering, the considered ladder diagrams for the irreducible polarizability $P$ (b), and the $T$-matrix
contribution $\Sigma^T$ to the self-energy (c). (d) Feynman diagrams for $K_{\sigma\sigma'}(Q)$ and
$M_{\sigma\sigma'}(Q)$. The wiggly lines signify the dynamically screened Coulomb interaction $W$. The solid
lines with arrows represent the Green function $G$. The $T$ matrix is shown by the shaded
square.}\label{K_M_ladder_diags}
\end{figure}

%++++++++++++++++++++++++++++++++++++++++++++++++++++++++++++++++++++++++++++++++++++++++++++++++++++++++
The $T$ matrix shown in Fig.~\ref{K_M_ladder_diags}(a) allows one to treat the problem of summation of
infinite classes of ladder diagrams in the diagrammatic expansion of both the irreducible polarizability
[Fig.~\ref{K_M_ladder_diags}(b)] and the quasiparticle self-energy [Fig.~\ref{K_M_ladder_diags}(c)]. In
momentum space, within the local approximation the $T$ matrix accounting for multiple electron-hole ($e-h$)
scattering has the form\cite{IAN_EVC}
\begin{equation}\label{TrialSolutionType}
\tilde{\Gamma}_{\sigma\sigma'}(Q)=\frac{\widetilde{W}_{\sigma\sigma'}(Q)}{1-\widetilde{W}_{\sigma\sigma'}(Q)K_{\sigma\sigma'}(Q)}.
\end{equation}
Here and in the following we use the four-momentum variables $Q$, $k$, or $p$ as a shorthand for
$(\mathbf{Q},\Omega)$, $(\mathbf{k},\omega)$, or $(\mathbf{p},\overline{\omega})$, respectively. $\sigma$
labels the spin, and $\sigma=\pm$ corresponds to spin-up, and spin-down, respectively. The $e-h$ propagator
$K_{\sigma\sigma'}$ is given by\cite{RPA_P0}
\begin{eqnarray}
&&K_{\sigma\sigma'}(Q)=\int dp\,\kappa_{\sigma \sigma',Q}(p),\label{K_big}\\
&&\kappa_{\sigma \sigma',Q}(p)=\frac{i}{(2\pi)^4}G_{\sigma}(Q+p)G_{\sigma'}(p),\label{kappa_def}
\end{eqnarray}
where $G_{\sigma}$ is the Green function. The local electron-hole interaction $\widetilde{W}_{\sigma\sigma'}$
is related with the dynamically screened Coulomb interaction $W$ by the equation [see
Fig.~\ref{K_M_ladder_diags}(d)]
\begin{equation}
\widetilde{W}_{\sigma\sigma'}(Q)=
[K_{\sigma\sigma'}(Q)]^{-1}M_{\sigma\sigma'}(Q)[K_{\sigma\sigma'}(Q)]^{-1},\label{local_interaction}
\end{equation}
where
\begin{equation}
M_{\sigma\sigma'}(Q)=\int dk\,dp\,\kappa_{\sigma \sigma',Q}(k)W(k-p)\kappa_{\sigma
\sigma',Q}(p).\label{first_exchange}
\end{equation}
%=============================================================================================================
%=============================================================================================================
Note that the $T$ matrix (\ref{TrialSolutionType}) depends only on the four-momentum transfer along the
electron-hole channel.

In spite of this simple form, a calculation of the $T$ matrix for the real system is still difficult due to
the eightfold integration in Eq.~(\ref{first_exchange}) defining the local interaction. In this paper, in
order to obtain a suitable expression for $\widetilde{W}_{\sigma\sigma'}$, we consider the small
four-momentum transfer limit (i.e., we set $\Omega=0$ and then take the limit $\mathbf{Q}\rightarrow0$). It
is expected to be a reasonable approximation because we are mainly interested in excitations in the vicinity
of the Fermi energy where quasiparticles are well defined.
%++++++++++++++++++++++++++++++++++++++++++++++++++++++++++++++++++++++++++++++++++++++++++++++++++++++++
\subsection{\label{sec_theory:local_interaction}Local interaction}
%++++++++++++++++++++++++++++++++++++++++++++++++++++++++++++++++++++++++++++++++++++++++++++++++++++++++
First of all we examine the spin-diagonal part of the local interaction. Omitting argumentations which one
can find in Ref.~\onlinecite{AGDz}, in the calculation of poles contribution to the integral
(\ref{first_exchange}) we replace $\kappa_{\sigma \sigma,Q}(k)$ by
$C_{\sigma\sigma}\delta(\omega)\delta[\epsilon_{\sigma}(\mathbf{k})]$. Unless stated otherwise, the
quasiparticle energy $\epsilon_{\sigma}(\mathbf{k})$ is measured from the Fermi energy $\epsilon_F$. The
coefficient $C_{\sigma\sigma}$ is determined by the relation (\ref{K_big}). Thus, $\kappa_{\sigma
\sigma,Q}(k)$ can be written as
\begin{equation}\label{kappa_approx}
\kappa_{\sigma \sigma,Q}(k)=\frac{1}{8\pi^3N^{\sigma}_F}K_{\sigma\sigma}(Q)
\delta(\omega)\delta[\epsilon_{\sigma}(\mathbf{k})]+\varphi(k),
\end{equation}
where $\varphi(k)$ is the regular part of $\kappa_{\sigma \sigma,Q}(k)$ and $N^{\sigma}_F$ is the density of
states of spin $\sigma$ per unit volume at the Fermi surface.\cite{remark_DOS} Substituting
Eq.~({\ref{kappa_approx}}) into Eq.~({\ref{first_exchange}}) and neglecting\cite{remark_RP} terms with
$\varphi$, we obtain that at the considered limit the local interaction has the form
\begin{equation}\label{LI_approx}
\widetilde{W}_{\sigma\sigma}=\frac{\int d\mathbf{k} d\mathbf{p} \delta[\epsilon_{\sigma}(\mathbf{k})]
W(\mathbf{k}-\mathbf{p},0)\delta[\epsilon_{\sigma}(\mathbf{p})]}{[8\pi^3N^{\sigma}_F]^2}
\end{equation}
or
\begin{equation}\label{LI_approx_FS}
\widetilde{W}_{\sigma\sigma}=\frac{1}{[8\pi^3N^{\sigma}_F]^2}
\int_{S_F^{\sigma}}\frac{dS_{\mathbf{k}}}{|\mathbf{v}_{\sigma}(\mathbf{k})|} W(\mathbf{k}-\mathbf{p},0)
\frac{dS_{\mathbf{p}}}{|\mathbf{v}_{\sigma}(\mathbf{p})|},
\end{equation}
where $S_F^{\sigma}$ is the Fermi surface for spin $\sigma$, and
$\mathbf{v}_{\sigma}(\mathbf{k})=\nabla_{\mathbf{k}}\epsilon_{\sigma}(\mathbf{k})$ is the quasiparticle
velocity at this surface. Now we have the fourfold integration instead of the eightfold one in Eq.~(\ref{first_exchange}). %Thus, at the

For the HEG in the spin-polarized state, Eq.~(\ref{LI_approx_FS}) can be rewritten as
\begin{equation}\label{LI_approx_HEG}
\widetilde{W}_{\sigma\sigma}=\frac{1}{(4\pi)^2}\int d\Omega _{\mathbf{k}} d\Omega _{\mathbf{p}}
W(k_F^{\sigma}|\hat{\mathbf{k}}-\hat{\mathbf{p}}|,0),
\end{equation}
where $\Omega _{\mathbf{k}}$ and $\Omega _{\mathbf{p}}$ are spatial angles, $\hat{\mathbf{k}}$ and
$\hat{\mathbf{p}}$ are unit vectors. The Fermi wave vector $k_F^{\sigma}$ for spin $\sigma$ is related to
that in the paramagnetic state $k_F$ by $k_F^{\sigma}=k_F(1-\sigma\zeta)^{1/3}$, where the relative spin
polarization $\zeta=|n_+-n_-|/n$ ($n_{\sigma}$ is the spin $\sigma$ electron density, $n=n_++n_-$ being the
HEG electron density) defines the exchange splitting of the band as
$2\Delta=\frac{k_F^2}{2}[(1+\zeta)^{2/3}-(1-\zeta)^{2/3}]$.\cite{Moriya} Note that considering the HEG, here
and in the following, in all expressions we use the noninteracting energy
$\epsilon_{\sigma}^0(\mathbf{k})=\mathbf{k}^2/2-k_F^{\sigma\,2}/2$ in place of
$\epsilon_{\sigma}(\mathbf{k})$, but for the quasiparticle energy in this case the notation $E_k$ is used
(see Sec.~\ref{sec_qp_properties:properties}).

As to the spin-non-diagonal part of the local interaction, instead of $\delta[\epsilon_{\sigma}(\mathbf{k})]$
in the expression replacing $\kappa_{\sigma\,-\sigma,Q}(k)$ we can use the ratio
\begin{equation}\label{gamma_ratio}
\gamma_{\sigma\,-\sigma}(\mathbf{k})=\frac{n_{F}[\epsilon_{\sigma}(\mathbf{k})]-n_{F}[\epsilon_{-\sigma}(\mathbf{k})]}
{\epsilon_{-\sigma}(\mathbf{k})-\epsilon_{\sigma}(\mathbf{k})},
\end{equation}
where $n_F$ is the Fermi distribution function. In this case, the coefficient $C_{\sigma\,-\sigma}$ is equal
to $K_{\sigma\,-\sigma}(Q)/[8\pi^3K_{\sigma\,-\sigma}(0)]$. As a result, the spin-non-diagonal local
interaction is given by the form
\begin{equation}\label{LI_approx_snd}
\widetilde{W}_{\sigma\,-\sigma}=\frac{\int d\mathbf{k} d\mathbf{p}
\gamma_{\sigma\,-\sigma}(\mathbf{k})W(\mathbf{k}-\mathbf{p},0)\gamma_{\sigma\,-\sigma}(\mathbf{p})}
{[8\pi^3K_{\sigma\,-\sigma}(0)]^2}.
\end{equation}

For the HEG in the spin-polarized state, we can rewrite Eq.~(\ref{LI_approx_snd}) in the following way
\begin{eqnarray}\label{LI_approx_snd_HEG}
\widetilde{W}_{\sigma\,-\sigma}&=&\frac{1}{(4\pi)^2}\int d\Omega _{\mathbf{k}} d\Omega _{\mathbf{p}}
\frac{1}{\left[2\zeta k_F^3/3\right]^2} \\
&\times& \int_{k_{F}^{\sigma}}^{k_{F}^{-\sigma}}|\mathbf{k}|^2d|\mathbf{k}|
\int_{k_{F}^{\sigma}}^{k_{F}^{-\sigma}}|\mathbf{p}|^2d|\mathbf{p}|W(\mathbf{k}-\mathbf{p},0).\nonumber
\end{eqnarray}

Note that the ratio $\gamma_{\sigma\,-\sigma}$ tends to the $\delta$-function at the $\zeta\rightarrow0$
limit (the paramagnetic state), and the spin-diagonal and spin-non-diagonal parts become equal. Since in this
paper we are interested in paramagnetic systems, the local interaction with $\zeta\neq0$ remains to be
examined elsewhere.

%++++++++++++++++++++++++++++++++++++++++++++++++++++++++++++++++++++++++++++++++++++++++++++++++++++++++
\subsection{\label{sec_theory:local_field_factor}Local-field factor}
%++++++++++++++++++++++++++++++++++++++++++++++++++++++++++++++++++++++++++++++++++++++++++++++++++++++++

In paper I, by examining the irreducible polarizability exchange diagrams [Fig.~\ref{K_M_ladder_diags}(b)],
we have shown that for paramagnetic systems the local electron-hole interaction $\widetilde{W}$ can be
identified with the local-field factor ${\cal G}(Q)$ (or more precisely with its exchange part)
\begin{equation}\label{LFF_EXCH}
{\cal G}(Q)=\widetilde{W}(Q) /2v_c(\mathbf{Q}),
\end{equation}
where $\widetilde{W}(Q)=\frac{1}{2}\sum_{\sigma}\widetilde{W}_{\sigma \sigma}(Q)$, and $v_c(\mathbf{Q})$ is
the bare Coulomb interaction. This exchange part is related to the first-order exchange diagram in the
irreducible polarizability diagrammatic expansion [see Eqs.~(\ref{local_interaction}), (\ref{first_exchange})
and Figs.~\ref{K_M_ladder_diags}(b) and \ref{K_M_ladder_diags} (d)]. Such a relation was derived and examined
by many authors within
various approaches (see, e.g., Refs.~\onlinecite{Richardson,LFF_Shturm, LFF_TokPan}, and \onlinecite{LFF_StubTokPan, %LFF_RORO,
LFF_Adragna,LFF_Marini,LFF_MBPT_TDDFT}). Thus we can verify the approximation done for $\widetilde{W}$ in the
previous subsection.

Actually, at the considered limit ($Q\rightarrow0$), the local-field factor of the HEG has %tends to
the form %given by
\begin{equation}\label{small_q_limit}
{\cal G}(Q)=A\left(\frac{\mathbf{Q}}{k_F}\right)^{2},
\end{equation}
where $A=\frac{\widetilde{W}k_F^2}{8\pi}$ in our case. This relation allows us to carry out, by the example
of the HEG in the paramagnetic state, a comparison of the coefficient $A$ obtained with the help of
Eq.~(\ref{LI_approx_HEG}) with that known from the literature.

\begin{figure}[tbp]
\centering
 \includegraphics[angle=0,scale=0.47]{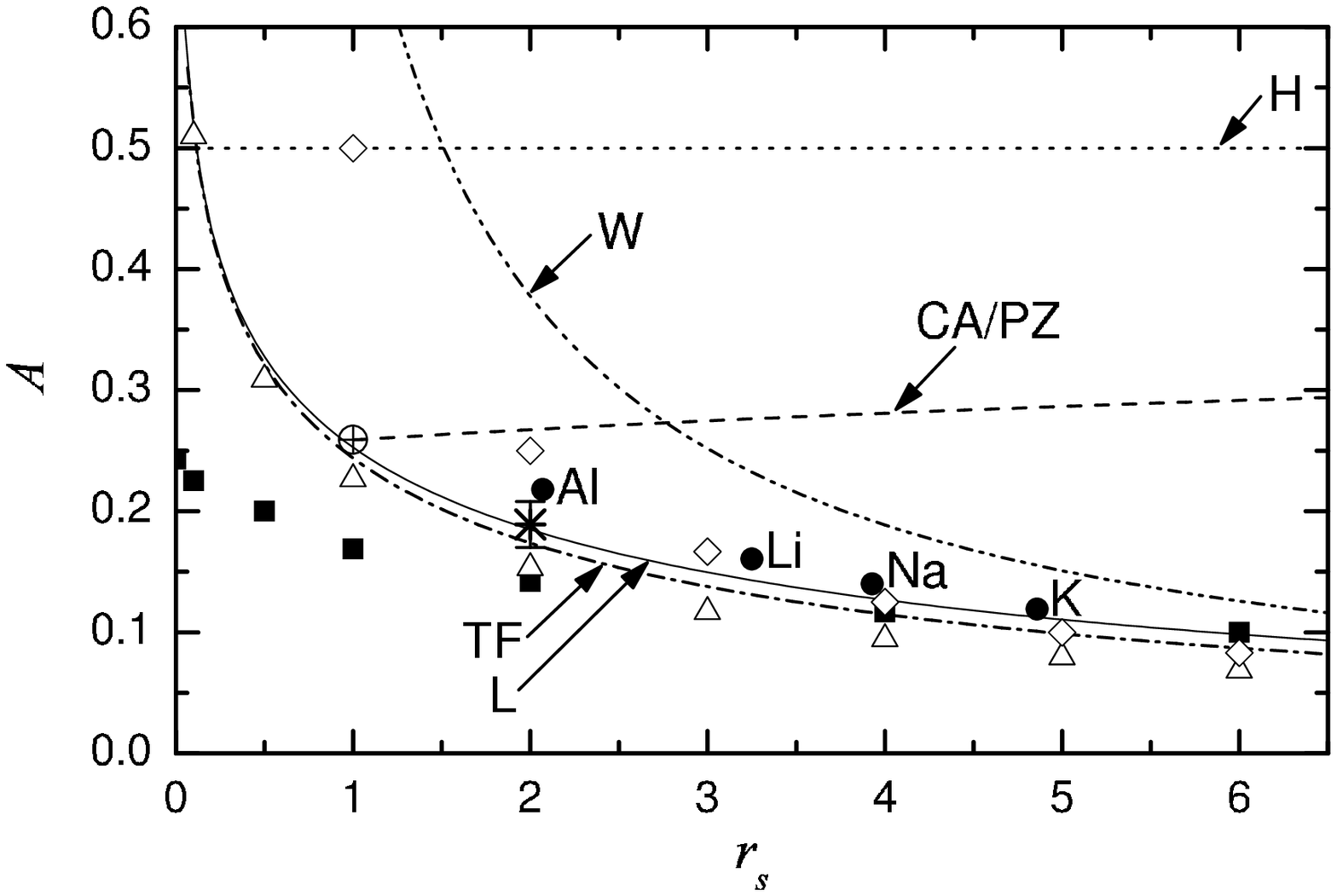}
\caption{The coefficient $A$ versus the electron density parameter $r_s$. Filled squares and circles
represent the results of Refs.~\onlinecite{Suehiro_Ousaka_Yasuhara} and \onlinecite{Tsolakidis},
respectively. Stars with the error bar depict $A$ obtained from calculations of ${\cal G}_a$ for $r_s=2$
shown in Fig.~3 of Ref.~\onlinecite{Richardson}. The notations H, CA/PZ, TF, L, and W signify, respectively,
the Hubbard approximation, the results of Monte Carlo calculations of Ref.~\onlinecite{CA_PZ} ($r_s\geq1$),
the Thomas-Fermi model [Eq.~(\ref{a_coeff_TF})], calculations by means of Eq.~(\ref{LI_approx_HEG}) with the
use of the Lindhard function, and, finally, $A$ obtained from Eq.~(\ref{low_den_lim}). Open triangles and
diamonds depict the ``self-consistent'' calculations (see the text) for the TF and W cases,
respectively.}\label{a_compress}
\end{figure}

The simplest way to analytically perform the integration in Eq.~(\ref{LI_approx_HEG}) is to use the screened
Coulomb interaction $W(\mathbf{k},0)=4\pi\left[|\mathbf{k}|^2+q_{TF}^2\right]^{-1}$ of the Thomas-Fermi model
(see, e.g., Ref.~\onlinecite{MahanBook}). Here $q_{TF}=\bigl{(}\frac{4\alpha r_s}{\pi}\bigr{)}^{1/2}k_F$ is
the Thomas-Fermi wave vector, the electron density parameter $r_s$ is given by
$\frac{4\pi}{3}(a_0r_s)^3=1/n$, $a_0$ being the Bohr radius, $\alpha=\left(\frac{4}{9\pi}\right)^{1/3}$, and
$\alpha r_s k_F=1$. As a result, for $A$ we obtain the following dependence on the electron density
parameter:
\begin{equation}\label{a_coeff_TF}
A_{TF}(r_s)=\frac{1}{8}\ln\left(1+\frac{\pi}{\alpha r_s}\right).
\end{equation}
In Fig.~\ref{a_compress}, we show $A_{TF}$ as a function of $r_s$ as well as the coefficient $A_L$ (the ``L''
curve) calculated by means of Eqs.~(\ref{LI_approx_HEG}), (\ref{LFF_EXCH}), and (\ref{small_q_limit}) with
the use of the Lindhard dielectric function. It follows from the figure that $A_{TF}$ and $A_L$ are very
close to each other, especially at $r_s\rightarrow0$. Moreover, both these results are in good agreement with
the coefficient $A$ obtained in Ref.~\onlinecite{Tsolakidis} (in the figure it is shown for Al, Li, Na, and
K), where the same class of ladder diagrams in the irreducible polarizability diagrammatic expansion was
considered.
%=============================================================================================================
%=============================================================================================================

Also, as shown in Fig.~\ref{a_compress}, our results are in accord with calculations performed in
Ref.~\onlinecite{Richardson} for the local-field factor ${\cal G}_a$ [see Eq.~(36) in the cited paper] which
is formally the same as that of Eq.~(\ref{LFF_EXCH}), except for including self-energy effects and the fact
that ${\cal G}_a$ is given on the imaginary-frequency axis. The shown error bar is originated from the
procedure of evaluating of $A$: inspecting Fig.~3 of Ref.~\onlinecite{Richardson}, we have found $A$ from a
set of values of ${\cal G}_a$ evaluated at various momenta ($0.4k_F$, $0.6k_F$, $0.8k_F$, and $1.0k_F$).

Due to the integration over the Fermi surface in Eq.~(\ref{LI_approx_HEG}), at the low-density limit
($r_s\rightarrow \infty$), when the ratio $q_{TF}/k_F$ is large, one can neglect momentum dependence of the
screened Coulomb interaction. This leads to $\widetilde{W}=W(0,0)$ and, as a consequence,
\begin{equation}\label{low_den_lim}
A_W(r_s)=\frac{\pi}{8\alpha r_s}.
\end{equation}
It is appropriate to mention here that the approximation made in Ref.~\onlinecite{Zhukov} corresponds to the
local interaction given by $\widetilde{W}(Q)=W(\mathbf{Q},0)$.  This means that the local-field factor of
Eq.~(\ref{LFF_EXCH}) can be represented as
\begin{equation}\label{inverse_epsilon_l}
{\cal G}(\mathbf{Q},0)=\frac{1}{2\varepsilon(\mathbf{Q},0)}.
\end{equation}
For the HEG, the approximation (\ref{inverse_epsilon_l}) gives a Hubbard-like form for ${\cal G}$ and at the
small four-momentum transfer limit it leads to $A=A_W$. Such an approximation can be valid at the low-density
limit, whereas for $r_s\lesssim2.5$ it essentially overestimates the exchange diagrams contribution,
exceeding even the accurate results of Monte Carlo calculations\cite{CA_PZ} which include both exchange and
correlation effects (see Fig.~\ref{a_compress}).

It is worth pointing out that the common property of $A_{TF}$, $A_{L}$, and $A_{W}$ is the divergence at
$r_s\rightarrow0$.\cite{remark_A_divergence} In this connection, it makes sense to compare our results with
the exchange local-field factor of Ref.~\onlinecite{Suehiro_Ousaka_Yasuhara} (filled squares in
Fig.~\ref{a_compress}). A class of diagrams considered there comprises ladder diagrams based on more complex
electron-hole interaction containing electron-electron and hole-hole multiple-scattering events. As a result,
the coefficient $A$ of Ref.~\onlinecite{Suehiro_Ousaka_Yasuhara} (evaluated from the local-field factor at
$k_F$ shown in Fig.~7 of the quoted paper) has no divergence in its dependence on $r_s$, tending to the
Hartree-Fock prediction $1/4$ at $r_s\rightarrow0$. Nevertheless, by inspecting Fig.~\ref{a_compress}, we can
infer that for the entire metallic density range ($r_s$ from 2 to 6) even the simplest approximation $A_{TF}$
yields results consistent with those known from the literature. This fact allows us to expect that the local
electron-hole interaction (\ref{LI_approx})--(\ref{LI_approx_HEG}) will give reasonable results on the
self-energy ladder diagrams treatment.

At last, we would like to note that in order to more precisely evaluate the exchange part of ${\cal G}$ from
Eqs.~(\ref{LI_approx_HEG}) and (\ref{LFF_EXCH}), one can include the corresponding local-field corrections
into $W$ and perform a kind of ``self-consistent'' ($sc$) procedure.\cite{remark_SC} As shown in
Fig.~\ref{a_compress} (open triangles and diamonds), such a procedure modifies $A_{TF}$ slightly, whereas in
the case of $A_W$ the changes are more significant and the resultant $A_{W:sc}=\frac{2}{3}A_W$ allows one to
reasonably estimate the considered diagrams contribution in the entire metallic density
range.\cite{remark_Hubbard} Moreover, the use of this prefactor $2/3$ together with the approximation
(\ref{inverse_epsilon_l}) in \textit{ab initio} calculations\cite{Moennich} brings theory and experiment to a
better agreement.

%++++++++++++++++++++++++++++++++++++++++++++++++++++++++++++++++++++++++++++++++++++++++++++++++++++++++
\section{\label{sec:qp_properties}Quasiparticle properties}
%++++++++++++++++++++++++++++++++++++++++++++++++++++++++++++++++++++++++++++++++++++++++++++++++++++++++
In this section, we address the question of how the electron-hole multiple scattering affects quasiparticle
properties which are determined by both the self-energy and its derivatives with respect to momentum and
frequency. We study the HEG in the paramagnetic state at different values of $r_s$ ranging from 2 to 56. This
study includes the metallic density range ($2\lesssim r_s\lesssim6$) as well as the range of large
$r_s\sim48$ for which the effective mass ``divergence'' was predicted in Ref.~\onlinecite{Zhang_et_al}.

%++++++++++++++++++++++++++++++++++++++++++++++++++++++++++++++++++++++++++++++++++++++++++++++++++++++++
\subsection{\label{sec_qp_properties:self-energy}Quasiparticle self-energy}
%++++++++++++++++++++++++++++++++++++++++++++++++++++++++++++++++++++++++++++++++++++++++++++++++++++++++
%=============================================================================================================
\begin{figure*}[tbp]
\centering
 \includegraphics[angle=0,scale=0.7]{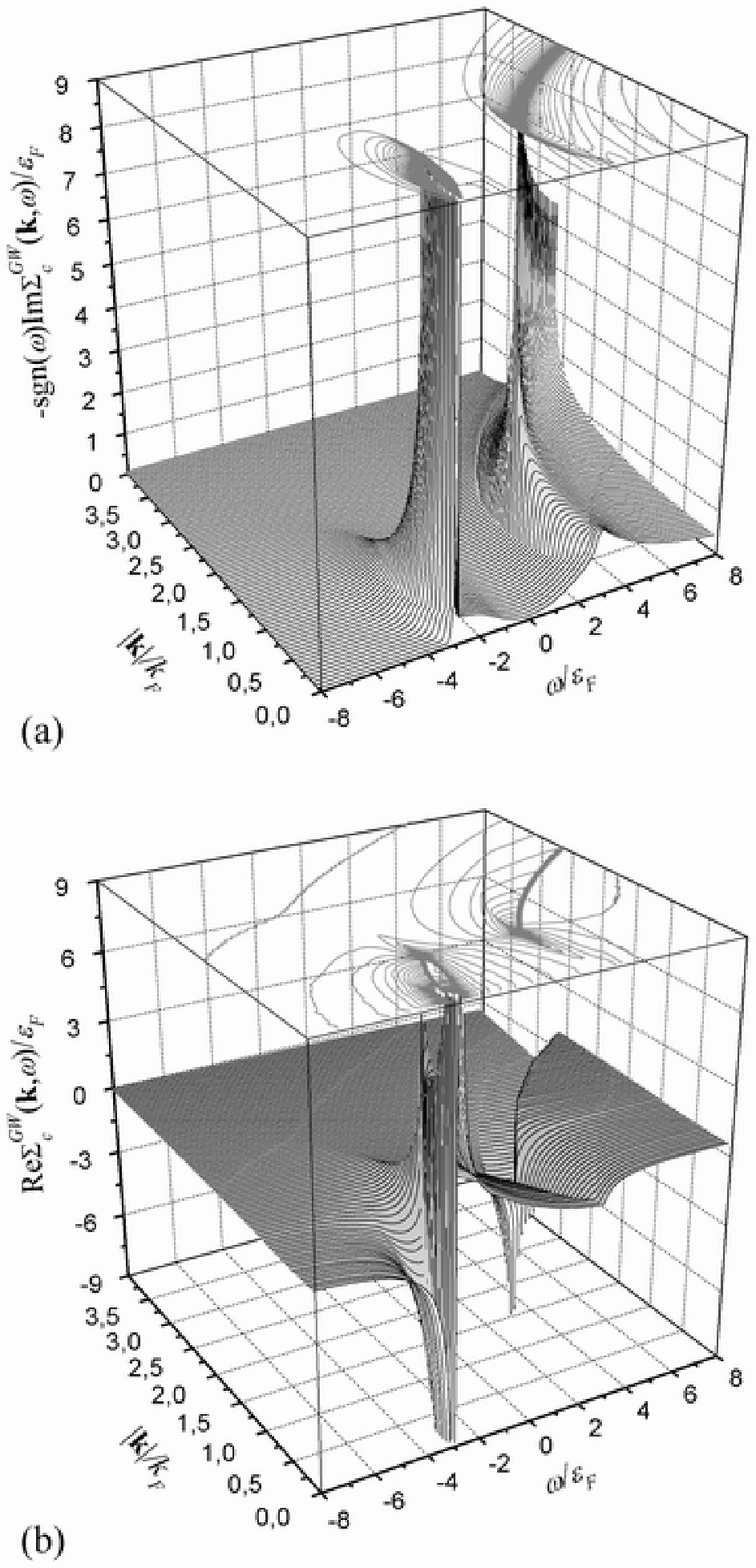}
 \includegraphics[angle=0,scale=0.7]{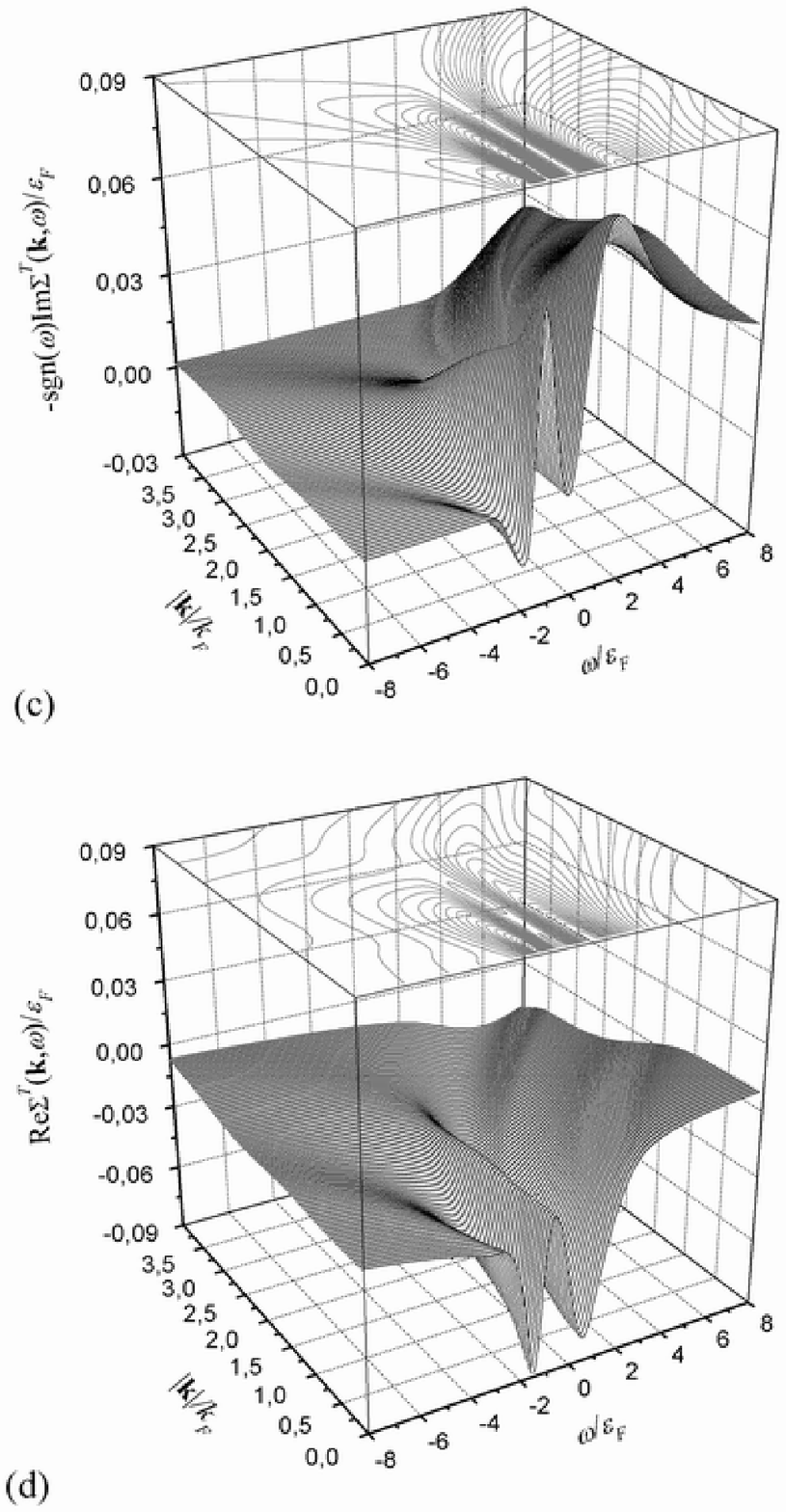}
\caption{The spectral function $-\mathrm{sgn}(\omega)\mathrm{Im}\Sigma_{c}^{GW}$ (a) and the real part
$\mathrm{Re}\Sigma_{c}^{GW}$ (b) of the correlation part of the self-energy calculated within the GWA. Also
the spectral function $-\mathrm{sgn}(\omega)\mathrm{Im}\Sigma^{T}$ (c) and the real part
$\mathrm{Re}\Sigma^{T}$ (d) of the $T$-matrix contribution to the self-energy. The GWA self-energy and the
$T$-matrix contribution are plotted as functions of energy $\omega/\epsilon_F$ and momentum
$|\mathbf{k}|/k_F$ for $r_s=4$. At the top of each of these figures, the contour plot (50 levels) is
shown.}\label{sigma3d}
\end{figure*}
%=============================================================================================================

In paper I, we have shown that the $T$ matrix (\ref{TrialSolutionType}) allows one to go beyond the GWA by
summing an infinite number of the electron-hole ladder diagrams shown in Fig.~\ref{K_M_ladder_diags}(c). In
this case, the quasiparticle self-energy can be expressed as
$\Sigma_{\sigma}=\Sigma^{GW}_{\sigma}+\Sigma^T_{\sigma}$, where the $GW$ term is well known to be
\begin{equation}\label{GW_Sigma}
\Sigma^{GW}_{\sigma}(p)=\frac{i}{(2\pi)^4}\int\,dkG_{\sigma}(k)W(p-k),
\end{equation}
and the $T$-matrix contribution is given by
\begin{equation}\label{MOMSTLocal}
\Sigma^{T}_{\sigma}(p)=-\frac{i}{(2\pi)^4}\sum_{\sigma'}\int\,dkG_{\sigma'}(k){\cal T}_{\sigma\sigma'}(p-k)
\end{equation}
with ${\cal T}_{\sigma\sigma'}(k)=\tilde{\Gamma}_{\sigma \sigma'}(k)[K_{\sigma
\sigma'}(k)\widetilde{W}_{\sigma\sigma'}(k)]^2$. It is important that the $T$-matrix contribution
(\ref{MOMSTLocal}) has a $GW$-like form that simplifies calculations of $\Sigma^T_{\sigma}$.

Due to the correspondence between multiple-scattering events and a spin fluctuation (see, e.g.,
Refs.~\onlinecite{Doniach_Engelsberg} and \onlinecite{Brinkman_Engelsberg, Hertz_Edwards, Riseborough, KIM,
Karlsson, Zhukov}), this approach to the self-energy includes the contributions of both charge and spin
fluctuations. As a result, in such an approach, having retained all the advantages of the $GW$ approximation,
we have a possibility of describing quasiparticle properties more comprehensively than it can be done in the
GWA. Note that as in the case of the irreducible polarizability ladder diagrams, the local interaction
$\widetilde{W}_{\sigma\sigma'}$ is an object of principal concern here because ${\cal T}_{\sigma\sigma'}$
depends significantly on the form for this interaction and, consequently, care must be taken by choosing an
approximation to it.

In this work, we mainly focus our attention on the effect of the inclusion of the self-energy ladder diagrams
on the quasiparticle properties in the case of the HEG in the paramagnetic state. For simplicity, we evaluate
the $GW$ term with the noninteracting Green function and the RPA screened interaction (the so-called $G_0W_0$
approximation which gives a better description of the quasiparticle properties than the fully self-consistent
$GW$ approximation itself, except for the total energy\cite{GV_electron_liquid,SC_GWA}). Below, unless stated
otherwise, this $G_0W_0$ approximation is referred to as the GWA. To evaluate the $T$-matrix contribution, we
also use the noninteracting Green function, the RPA electron-hole propagator, and the local interaction
elaborated in the previous section (corresponding to $A_L$).

For the computational purposes, it is convenient to split up the $GW$ term into the energy-independent
Hartree-Fock (exchange) part $\Sigma^{HF}(\mathbf{k})$ and the correlation part
$\Sigma_{c}^{GW}(\mathbf{k},\omega)$ defined through Eq.~(\ref{GW_Sigma}) by the induced potential
$W_{i}=W-v_c$ instead of the screened interaction $W$. Within the spectral function
representation,\cite{FerdiAnis} having found the imaginary part of $\Sigma_{c}^{GW}$ we obtain its real part,
$\mathrm{Re}\Sigma_c^{GW}(\mathbf{k},\omega)$, from the Hilbert transform by using the principal value
integration. As a result, the real part of the self energy in the GWA is equal to
$\mathrm{Re}\Sigma^{GW}(\mathbf{k},\omega)=\Sigma^{HF}(\mathbf{k})+\mathrm{Re}\Sigma_c^{GW}(\mathbf{k},\omega)$.
Regarding the $T$-matrix contribution, we first evaluate $\mathrm{Im}\Sigma^{T}$ of Eq.~(\ref{MOMSTLocal})
and then perform the Hilbert transform.

In order to get some idea of the quantity and behavior of the $T$-matrix contribution in comparison with the
$GW$ term, in Fig.~\ref{sigma3d} we plot the correlation part of the self-energy $\Sigma_{c}^{GW}$ and %as well as
the $T$-matrix contribution $\Sigma^{T}$ as a function of the momentum and energy for $r_s=4$. As follows
from the figure, due to plasmon singularities of the inverse dielectric function, the $GW$ term seems
substantially larger than the $T$-matrix contribution.\cite{remark_magnon} However, inspecting
Figs.~\ref{sigma3d}(c) and \ref{sigma3d}(d), one can see that the $\Sigma^{T}$ surface becomes notably
corrugated in the vicinity of the $\omega=0$ energy where $\Sigma^{GW}$ behaves smoothly and contains only
the electron-hole contribution (the plasmon decay channel is not open yet in this energy range) which is
comparatively not large. As a consequence, the $T$-matrix inclusion is expected to appreciably influence the
quasiparticle properties (especially on those determined by the self-energy derivatives) in this energy
region.

%=============================================================================================================
\begin{figure}[tbp]
\centering
 \includegraphics[angle=0,scale=0.47]{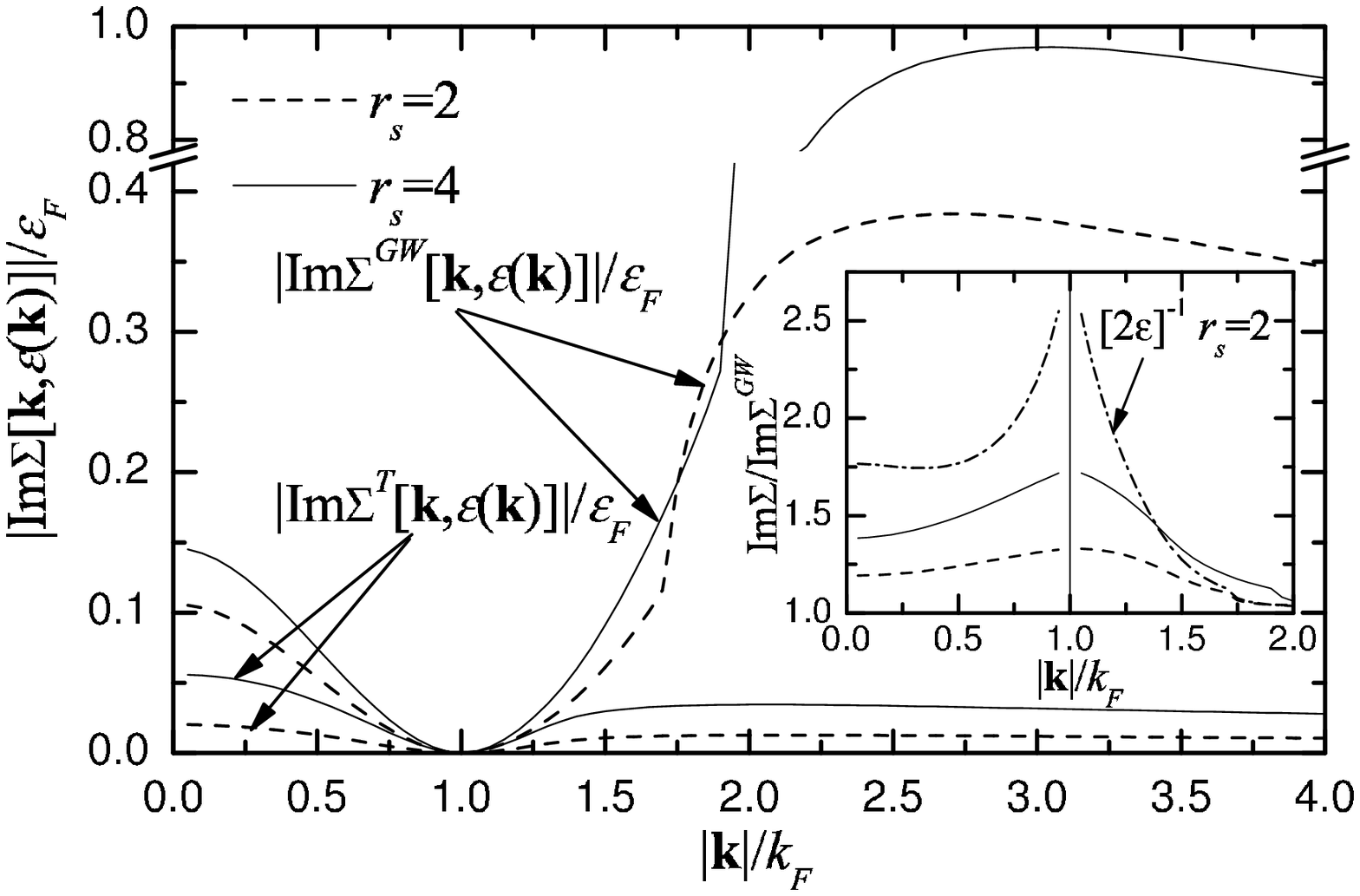} %scale=0.43
\caption{The absolute value of the imaginary part of the self-energy %$\mathrm{Im}\Sigma[\mathbf{k},\epsilon(\mathbf{k})]$
evaluated at $\omega=\epsilon(\mathbf{k})$ (the $GW$ term and the $T$-matrix contribution) as a function of
momentum $|\mathbf{k}|/k_F$ for $r_s=2$ (dashed line) and $4$ (solid line). Inset: the ratio
$\mathrm{Im}\Sigma/\mathrm{Im}\Sigma^{GW}=1+\mathrm{Im}\Sigma^{T}/\mathrm{Im}\Sigma^{GW}$ as a function of
momentum $|\mathbf{k}|/k_F$ for $r_s=2$ and $4$. The dash-dotted line represents
this ratio evaluated with the local-field factor of Eq.~(\ref{inverse_epsilon_l}) for $r_s=2$.
}\label{im_sigma_2_4}
\end{figure}
%=============================================================================================================
%=============================================================================================================
\begin{figure}[tbp]
\centering
 \includegraphics[angle=0,scale=0.47]{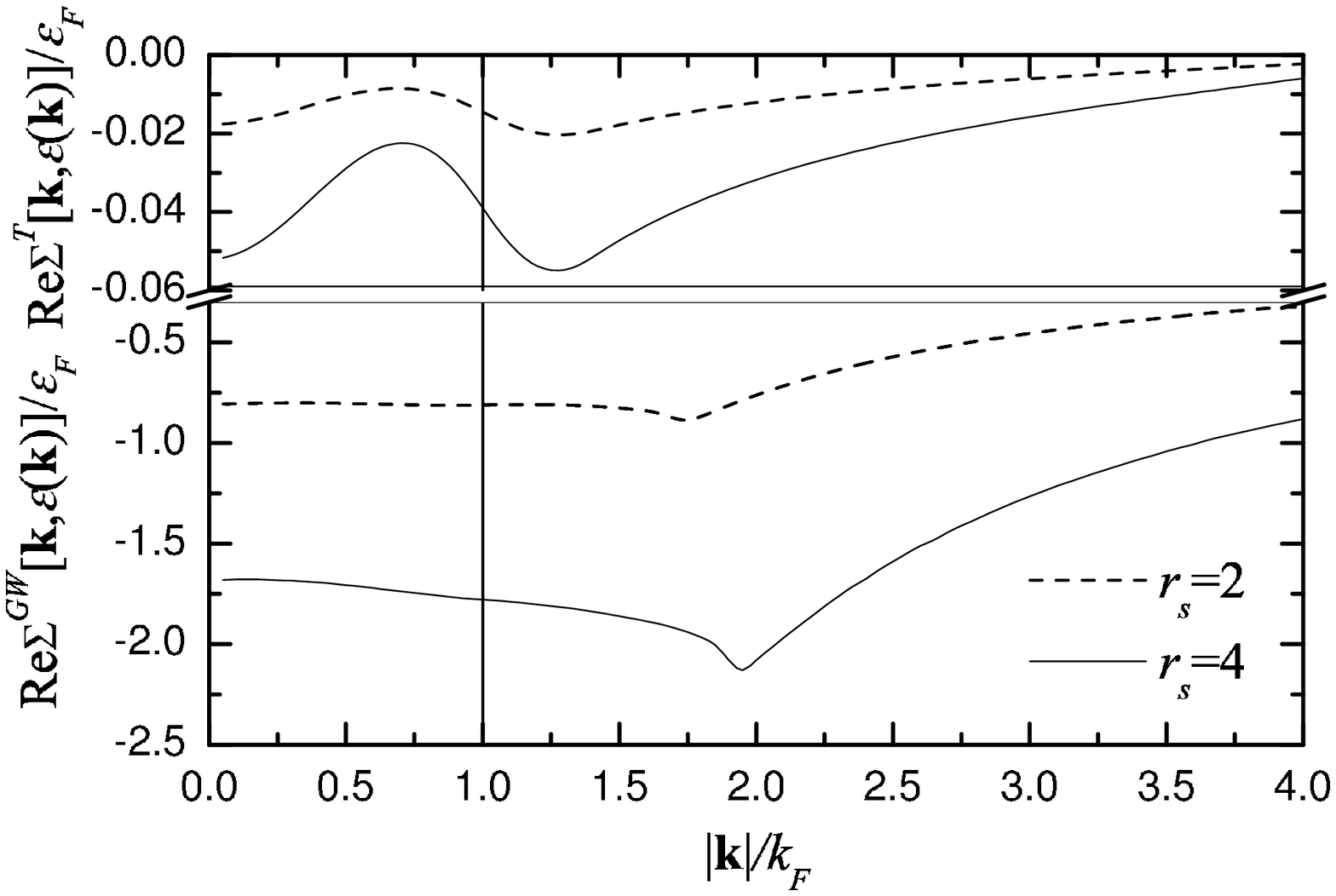} %scale=0.43
\caption{The real part of the self-energy $\mathrm{Re}\Sigma[|\mathbf{k}|,\epsilon(\mathbf{k})]$ (the $GW$ term
and the $T$-matrix contribution) as a function of momentum $|\mathbf{k}|/k_F$ for $r_s=2$ (dashed line) and
$r_s=4$ (solid line).}\label{re_sigma_2_4}
\end{figure}
%=============================================================================================================
As is evident from Fig.~\ref{sigma3d}(c), at some energies $\omega<0$ the spectral function of the $T$-matrix
contribution becomes negative. However, it does not lead to wrong analytical properties of the self-energy,
because the contribution (\ref{MOMSTLocal}) has such a form as an additional term beyond the $GW$ one and,
consequently, should be considered in sum with the latter. This sum is always non-negative.

Now we consider the on-shell imaginary part of the self-energy which gives the quasiparticle damping rate. In
Fig.~\ref{im_sigma_2_4}, we show this quantity as a function of momentum $|\mathbf{k}|/k_F$ for $r_s=2$ and
$4$. First of all, we would like to note that despite the fact that the spectral function of
$\Sigma^{T}(\mathbf{k},\omega)$ is negative within some $\mathbf{k}-\omega$ domain the on-shell imaginary
part of the $T$-matrix contribution has the same sign as the $GW$ term at any value of momentum. This means
that the $T$-matrix inclusion leads to an increase of the quasiparticle damping rate with respect to that in
the GWA. In order to assess the quantity of this increase, we examine the ratio
$\mathrm{Im}\Sigma/\mathrm{Im}\Sigma^{GW}=1+\mathrm{Im}\Sigma^{T}/\mathrm{Im}\Sigma^{GW}$ (see inset of
Fig.~\ref{im_sigma_2_4}). According to the calculations performed, this ratio as a function of momentum has
the largest value at $|\mathbf{k}|\sim k_F$ and falls down when $|\mathbf{k}|$ moves away from $k_F$.
Furthermore, the lower the electron density that the HEG possesses, the larger the ratio we have. For
example, $\mathrm{Im}\Sigma/\mathrm{Im}\Sigma^{GW}$ evaluated at $|\mathbf{k}|=1.05k_F$ is equal to $1.33$
for $r_s=2$ and $1.72$ for $r_s=4$.

As compared with the generalized $GW$ self-energy evaluated in Ref.~\onlinecite{GV_electron_liquid}, for
$|\mathbf{k}|<k_F$ we have similar results which are very close quantitatively. However, it is not so for
$|\mathbf{k}|>k_F$ especially in the region where a quasiparticle can decay into plasmons. Under this region
the on-shell imaginary part of the generalized $GW$ self-energy very quickly becomes smaller than that in the
GWA and then demonstrates a partial ``suppression'' of the plasmon decay channel in comparison with the $GW$
approximation (see also Ref.~\onlinecite{Santoro_Giuliani}).

Next, to answer the question of how the value of the ratio $\mathrm{Im}\Sigma/\mathrm{Im}\Sigma^{GW}$ depends
on the chosen form of the approximation for the local electron-hole interaction, we have calculated
$\mathrm{Im}\Sigma^{T}$ with $\widetilde{W}$ defined by Eq.~(\ref{LFF_EXCH}) with the local-field factor
(\ref{inverse_epsilon_l}) for $r_s=2$. As follows from the inset of Fig.~\ref{im_sigma_2_4}, the ratio
becomes essentially larger than that presented by dashed line and runs up to $2.53$ at
$|\mathbf{k}|=1.05k_F$. This means that the use of such an interaction $\widetilde{W}$ instead of
Eq.~(\ref{LI_approx_HEG}) leads to an increase of the ratio by a factor of $1.90$ at this momentum. Thus,
keeping in mind that at the small four-momentum transfer limit this local interaction corresponds to the
coefficient $A_W$ of Eq.~(\ref{low_den_lim}), we suppose that at least for aluminum the use of this
approximation to $\widetilde{W}$ results in a large value of the $T$ matrix that in turn rises the
quasiparticle damping rate too high. In connection with the coefficients shown in Fig.~\ref{a_compress}, we
would like to note that, e.g., for $r_s=2$ the local-field factor of Ref.~\onlinecite{Tsolakidis} used in
paper I gives the increase of the ratio by a factor of $1.24$.

Regarding the on-shell real part of the self-energy (see Fig.~\ref{re_sigma_2_4}), we would like to emphasize
that again the $T$-matrix contribution has the same sign as the $GW$ term. Owing to the ``corrugations''
discussed in connection with Fig.~\ref{sigma3d}, in the vicinity of $k_F$ $\mathrm{Re}\Sigma^{T}$ shows fast
variation which for $0.7k_F\lesssim|\mathbf{k}|\lesssim1.3k_F$ can be fitted rather well by a sinelike
function. As follows from Fig.~\ref{re_sigma_2_4}, the $T$-matrix contribution represents itself as a very
small quantity within all the considered $\mathbf{k}$-domain and therefore can alter the on-shell real part
of the self-energy evaluated in the GWA to only a small extent.
%=============================================================================================================
\begin{figure}[tbp]
\centering
 \includegraphics[angle=0,scale=0.47]{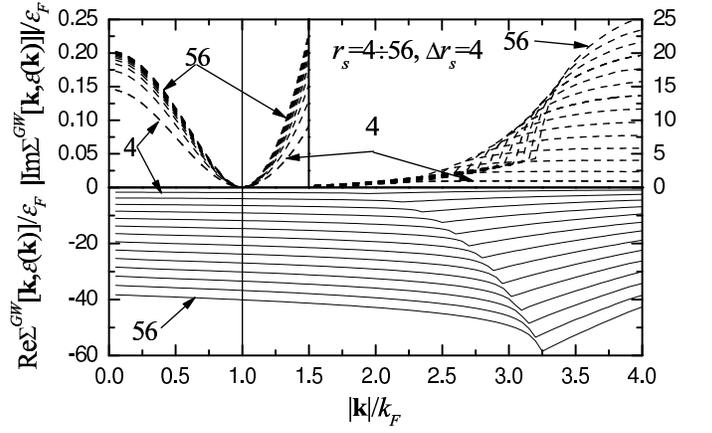} %scale=0.43
\caption{The on-shell real and imaginary parts of the self-energy calculated in the GWA as functions of
momentum $|\mathbf{k}|/k_F$ for $r_s$ values ranging from $4$ to $56$.}\label{sigma_gw_4_56}
\end{figure}
%=============================================================================================================
%=============================================================================================================
\begin{figure}[tbp]
\centering
 \includegraphics[angle=0,scale=0.47]{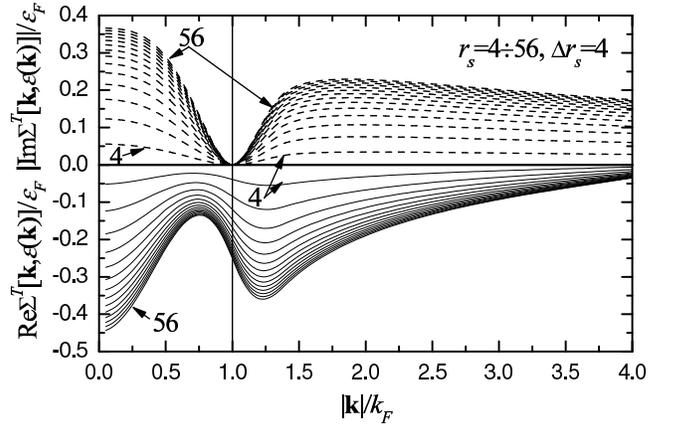} %scale=0.43
\caption{The on-shell real and imaginary parts of the $T$-matrix contribution as functions of momentum
$|\mathbf{k}|/k_F$ for $r_s$ values ranging from $4$ to $56$.}\label{sigma_tm_4_56}
\end{figure}
%=============================================================================================================

In Figs.~\ref{sigma_gw_4_56} and \ref{sigma_tm_4_56}, we show how the $GW$ term and the $T$-matrix
contribution depend on $r_s$ beyond the metallic density range. One can see that the on-shell
$\mathrm{Im}\Sigma^{GW}$ as a function of $r_s$ shows relatively small changes for
$0.0k_F\leq|\mathbf{k}|\lesssim1.5k_F$ and already at $r_s\sim24$ gets some ``saturation'', whereupon
properties of the $e-h$ decay channel remain practically unchanged within the mentioned $\mathbf{k}$
interval. For $|\mathbf{k}|>1.5k_F$, especially in the plasmon emission region, the quasiparticle damping
rate demonstrates a continual increase with increasing $r_s$. Owing to this and the $r_s$-dependence of
$\Sigma^{HF}$, the on-shell real part of the $GW$ self-energy monotonically decreases as a function of $r_s$.
As to the $T$-matrix contribution, after $r_s\sim40$ the changes of the imaginary and real parts become
insignificant for any momentum.\cite{remark_ferro}

\begin{figure}[tbp]
\centering
 \includegraphics[angle=0,scale=0.47]{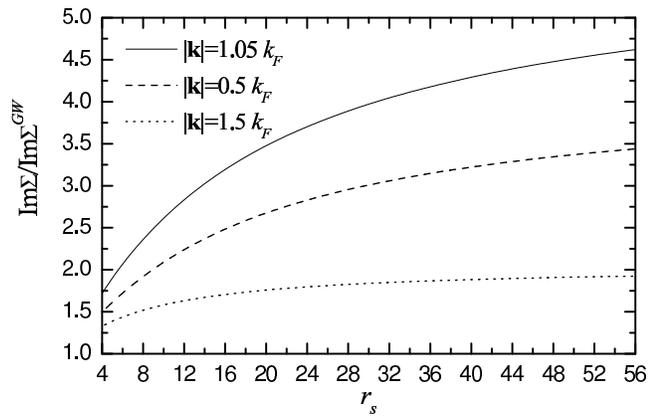} %scale=0.43
\caption{The ratio $\mathrm{Im}\Sigma/\mathrm{Im}\Sigma^{GW}=1+\mathrm{Im}\Sigma^{T}/\mathrm{Im}\Sigma^{GW}$
as a function of $r_s$ at $|\mathbf{k}|=0.5k_F$ (dashed line), $1.05k_F$ (solid line), and $1.5k_F$ (dotted
line).}\label{t_matrix_ratio_4_56}
\end{figure}

In Fig.~\ref{t_matrix_ratio_4_56}, we plot the ratio $\mathrm{Im}\Sigma/\mathrm{Im}\Sigma^{GW}$ as a function
of $r_s$ evaluated for different momenta. In fact this figure shows how the inclusion of the self-energy
ladder diagrams affects the quasiparticle damping rate. The biggest contribution of these diagrams is
observed in the vicinity of $k_F$ and can exceed the GWA prediction, e.g., by a factor of $3.6$ at $r_s=56$.

%=============================================================================================================
%=============================================================================================================

%++++++++++++++++++++++++++++++++++++++++++++++++++++++++++++++++++++++++++++++++++++++++++++++++++++++++
\subsection{\label{sec_qp_properties:properties}Quasipartical energy, renormalization constant,
and effective mass enhancement}
%++++++++++++++++++++++++++++++++++++++++++++++++++++++++++++++++++++++++++++++++++++++++++++++++++++++++

\begin{figure}[tbp]
\centering
 \includegraphics[angle=0,scale=0.47]{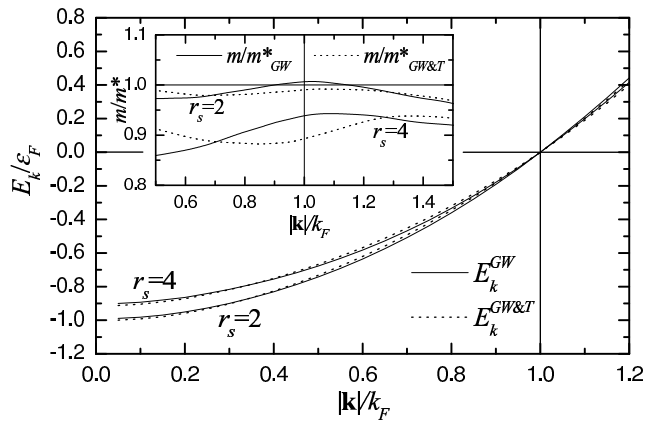} %scale=0.43
\caption{The quasiparticle energy $E_k$, Eq.~(\ref{qpenergy}), calculated without ($GW$, solid line) and with
($GW$\&$T$, dotted line) the $T$-matrix contribution as a function of momentum $|\mathbf{k}|/k_F$ for $r_s=2$
and $r_s=4$. Inset: the inverse effective mass enhancement $m/m^*$ as a function of $|\mathbf{k}|/k_F$ for
$r_s=2$ and $4$. }\label{dispersion_eff_mass}
\end{figure}

The quasiparticle excitation energy within the first-order perturbation theory is given by
\begin{equation}\label{qpenergy}
E_k=\epsilon(\mathbf{k})+\mathrm{Re}\Sigma[\mathbf{k},\epsilon(\mathbf{k})]-\mathrm{Re}\Sigma[k_F,0],
\end{equation}
where the real part of the self-energy in the on-shell approximation is added to the single-particle energy.
In Fig.~\ref{dispersion_eff_mass}, we show the calculated $E_k$ for $r_s=2$ and $4$. Due to the small value
of the on-shell real part of the $T$-matrix contribution, the quasiparticle dispersion is only slightly
changed by the inclusion of the self-energy ladder diagrams. Inspecting the figure, one can see that this
inclusion results in a light increase of the bandwidth with respect to the $GW$ one. This increase is similar
to that evaluated in Ref.~\onlinecite{MahanGWGamma} within the $GW\Lambda$ approximation with the inclusion
of the same vertex function in the screened interaction and the numerator of the self-energy. However this
similarity is observed only under $r_s\sim4$. For $r_s\gtrsim4$, the $GW\Lambda$ approximation yields band
narrowing greater than the GWA does.
%=============================================================================================================
%=============================================================================================================

As was anticipated, we have a more profound effect of the $T$-matrix inclusion on the quasiparticle
properties determined by the derivatives of the self-energy. One of such properties is the effective mass
enhancement which in the on-shell approximation is known to be given by
\begin{equation}\label{eme_osa}
\frac{m^*(\mathbf{k})}{m}=\left[\frac{m}{k}\frac{dE_k}{dk}\right]^{-1}.
\end{equation}
The inset of Fig.~\ref{dispersion_eff_mass} represents the inverse value of this dispersing quasiparticle
effective mass for $r_s=2$ and $4$. A close examination of $m/m^{*}$ as a function of momentum provides
interesting insights. First, owing to the sinelike behavior of the on-shell $\mathrm{Re}\Sigma^{T}$ in the
vicinity of $k_F$ (see Fig.~\ref{re_sigma_2_4}), the $T$-matrix inclusion gives alternating contribution to
the inverse effective mass. As a result, $m/m^{*}$ becomes smaller at $|\mathbf{k}|\sim k_F$ and greater away
from it. At some momenta the $T$-matrix contribution to the effective mass is equal to zero. Second, for
$r_s=2$ multiple electron-hole scattering modifies the effective mass to an extent that quasiparticles become
``heavier'' than in the noninteracting system, whereas the GWA predicts the reverse. Third, as can be seen
from Fig.~\ref{re_sigma_2_4}, for $r_s=4$ the absolute value of the derivative of the on-shell
$\mathrm{Re}\Sigma^{T}$ with respect to momentum should be larger than that for $r_s=2$. This entails the
larger alteration of $m/m^{*}$. Thus, focusing our attention on the effective mass behavior in the vicinity
of $k_F$, we can infer that the multiple electron-hole scattering leads to a ``weighting'' of quasiparticles.

%=============================================================================================================
\begin{figure}[tbp]
\centering
 \includegraphics[angle=0,scale=0.47]{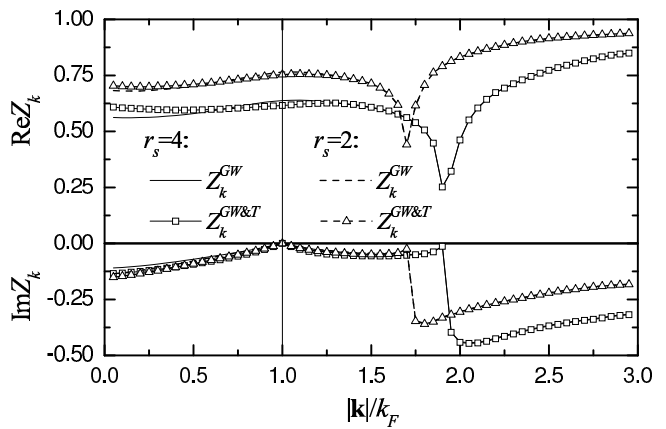} %scale=0.43
\caption{The real and imaginary parts of the renormalization constant $Z_k$ calculated without ($GW$) and
with ($GW$\&$T$) the $T$-matrix contribution as a function of momentum $|\mathbf{k}|/k_F$ for $r_s=2$ and
$4$.}\label{z_factor}
\end{figure}
%=============================================================================================================

%=============================================================================================================
\begin{table}[tbp]
\caption{\label{tab_table1} The effective mass enhancement $m^*/m$ and the renormalization constant $Z_F$ at
the Fermi wave vector $k_F$ calculated without ($GW$) and with ($GW$\&$T$) the $T$-matrix contribution. In
each of these cases, two values of the mass enhancement are presented: the first one is obtained from the
exact formula, Eq.~(\ref{eme_exact}), the second one (in parentheses) is calculated within the on-shell
approximation, Eq.~(\ref{eme_osa}).}
\begin{ruledtabular}
\begin{tabular}{ccccc}
 &\multicolumn{2}{c}{$m^*/m$}&\multicolumn{2}{c}{$Z_F$}\\
 $r_s$&$GW$&$GW$\&$T$&$GW$&$GW$\&$T$\\ \hline
 2&$0.99\, (0.99)$&$1.01\, (1.01)$&$0.77$&$0.75$ \\
 4&$1.05\, (1.07)$&$1.08\, (1.12)$&$0.64$&$0.62$\\
\end{tabular}
\end{ruledtabular}
\end{table}
%=============================================================================================================

Note that Eq.~(\ref{eme_osa}) is a valid approximation to the effective mass enhancement at small values of
$r_s$.\cite{GV_electron_liquid} In order to estimate the quasiparticle effective mass at $k_F$ more
precisely, especially for large $r_s$, we use the formally exact equation\cite{MahanBook}
\begin{equation}\label{eme_exact}
\frac{m^*}{m}=\frac{Z_F^{-1}}{1+\frac{m}{k_F}\left.\frac{\partial\Sigma(\mathbf{k},\omega)}{\partial k}
\right|_{\omega=0, |\mathbf{k}|=k_F}}.
\end{equation}
Here $Z_F$ is the renormlization constant $Z_k$ evaluated at the Fermi wave vector. In its turn, the
renormalization constant $Z_k$ that gives the spectral weight of the quasiparticle is defined for a
four-momentum $\left[\mathbf{k},\,\epsilon(\mathbf{k})\right]$ as\cite{MahanBook}
\begin{equation}\label{renorm_constant}
Z_k=\left[1-\left.\frac{\partial\Sigma(\mathbf{k},\omega)}{\partial\omega}
\right|_{\omega=\epsilon(\mathbf{k})}\right]^{-1}.
\end{equation}
%=============================================================================================================
\begin{figure}[tbp]
\centering
 \includegraphics[angle=0,scale=0.47]{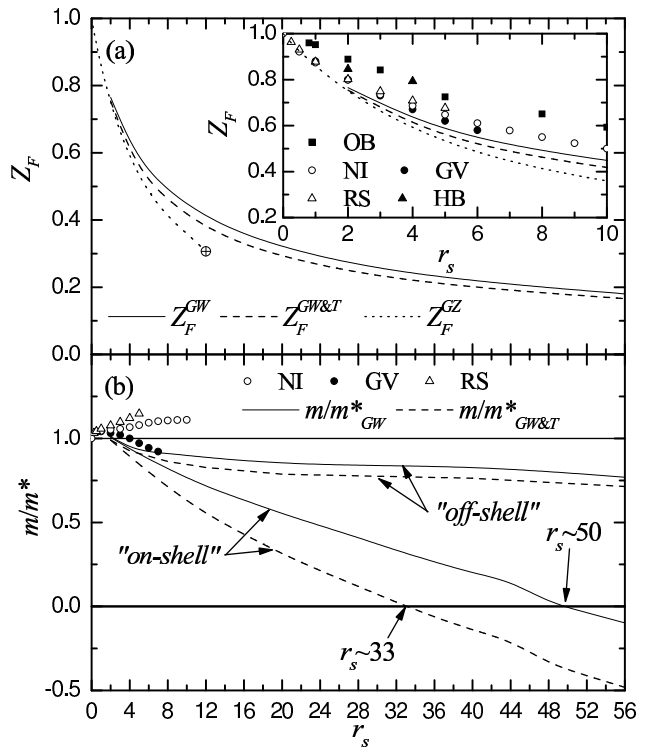} %scale=0.43
\caption{(a) The renormalization constant $Z_F$ at the Fermi wave vector obtained without ($GW$, solid line)
and with ($GW$\&$T$, dashed line) the $T$-matrix contribution as a function of $r_s$. The dotted line
represents the so-called GZ parametrization (Ref.~\onlinecite{Gori_Ziesche}) which is valid in the density
range $r_s\lesssim12$. The inset shows the obtained results in comparison with those known from the
literature. (b) The inverse effective mass enhancement $m/m^*$ at the Fermi wave vector calculated both in
the on-shell approximation, Eq.~(\ref{eme_osa}), labeled as ``on-shell'', and by making use of the formally
exact Dyson equation (\ref{eme_exact}), labeled as ``off-shell''. The notations OB, NI, GV, RS, and HB
signify the correspondent values taken from Refs.~\onlinecite{Ortiz_Ballone}, \onlinecite{Nakano_Ichimaru},
\onlinecite{GV_electron_liquid}, \onlinecite{Rietschel_Sham}, and \onlinecite{SC_GWA},
respectively.}\label{invers_eff_mass_Z_2_56}
\end{figure}
%=============================================================================================================

In Fig.~\ref{z_factor}, we plot the renormalization constant as a function of momentum for $r_s=2$ and $4$.
It follows from the figure that changes of the real part of $Z_k$ caused by the $T$-matrix inclusion occur
mainly up to $|\mathbf{k}|\sim 1.5k_F$. The imaginary part of $Z_k$ accounting for multiple electron-hole
scattering differs weakly from that in the GWA. At $|\mathbf{k}|=k_F$, the renormalization constant is real.
Due to the fact that the frequency derivative
$\left.\frac{\partial\mathrm{Re}\Sigma^{T}}{\partial\omega}\right|_{|\mathbf{k}|=k_F,\omega=0}$ is negative
as well as in the case of the $GW$ term (see Fig.~\ref{sigma3d}), $Z_F$ becomes smaller with the $T$-matrix
inclusion. Together with the positive momentum derivative of $\mathrm{Re}\Sigma^{T}$ it affects the effective
mass of Eq.~(\ref{eme_exact}) and moves the latter away from its value obtained in the GWA.
Table~\ref{tab_table1} contains our detailed results on the effective mass enhancement and the
renormalization constant for $r_s=2$ and $4$.

In Fig.~\ref{invers_eff_mass_Z_2_56}, we show the calculated dependence of $Z_F$ and $m/m^{*}$ on $r_s$ and
compare it with that known from the literature. As follows from Fig.~\ref{invers_eff_mass_Z_2_56}(a), the
inclusion of the $T$ matrix reduces the renormalization constant $Z_F$. Thereby the $T$-matrix contribution
shifts the $Z_F$-curve towards the so-called GZ (Gori-Giorgi--Ziesche) parametrization of
Ref.~\onlinecite{Gori_Ziesche}. This parametrization is in good agreement with calculations of
Ref.~\onlinecite{Takada_2} performed by making use of the effective-potential-expansion (EPX) method which in
contrast to the Hedin expansion\cite{Hedin} is formulated in terms of the static screened interaction
$W(\mathbf{q},0)$. The GZ parametrization is also compatible with QMC data found in
Ref.~\onlinecite{Ortiz_Ballone} for the HEG momentum distribution. The fact that the renormalization constant
$Z_F$ determined by Monte Carlo\cite{Ortiz_Ballone} (labeled as OB in the figure) is noticeably larger than
the parametrization can be explained by the difference in procedures of finding the momentum distribution
discontinuity at $k_F$. In the QMC calculations there is a finite distance between momenta (including the
closest to $k_F$) for which the momentum distribution is calculated (see, e.g., Fig.~12 in
Ref.~\onlinecite{Ortiz_Ballone}), whereas the GZ parametrization allows one to find $Z_F$ from
$|\mathbf{k}|\rightarrow k_F\pm0$ limits.

Note that the self-consistent schemes of Refs.~\onlinecite{SC_GWA} (HB), \onlinecite{Nakano_Ichimaru} (NI),
and \onlinecite{Rietschel_Sham} (RS), exhibit an increase of the quasiparticle renormalization constant at
the Fermi surface with respect to $Z_F^{GW}$ shown in Fig.~\ref{invers_eff_mass_Z_2_56}(a). It can be
understood within the framework of the detailed analysis carried out in Ref.~\onlinecite{SC_GWA}, where the
self-consistency between the Green function and the self-energy within the $GW$ approximation has been
achieved. As was shown,\cite{SC_GWA} the self-consistency procedure leads to the strongly suppressed plasmon
peaks in the imaginary part of the self-energy. Owing to the Hilbert transform, it entails an essential
smoothing of the corresponding sharp structures of the self-energy real part. As a consequence, the frequency
derivative of the latter, which is negative, becomes smaller by absolute value, that, in its turn, through
Eq.~(\ref{renorm_constant}) leads to an increase of the renormalization constant.

As regards the effective mass enhancement, it is evident from Fig.~\ref{invers_eff_mass_Z_2_56}(b) that the
on-shell approximation (\ref{eme_osa}) underestimates $m/m^{*}$ in comparison with that obtained from
Eq.~(\ref{eme_exact}). At large values of $r_s$ it comes into particular prominence. The on-shell $GW$
effective mass exhibits a divergence at $r_s\sim50$ (in Ref.~\onlinecite{Zhang_et_al} at $r_s\sim48$). The
$T$-matrix inclusion results in a considerable increase of the on-shell effective mass and leads to the
divergence at $r_s\sim33$. The effect of taking into account multiple electron-hole scattering on the
effective mass calculated by making use of the formally exact equation (\ref{eme_exact}) is consistent with
the influence of the $T$-matrix inclusion on the renormalization constant. In this case the quasiparticle
mass demonstrates a relatively weak $r_s$ dependence without any divergence up to the largest $r_s$
considered in the paper.

Regarding the $r_s$ dependence of the quasiparticle effective mass, one can note that the effective mass is a
more controversial quantity than the renormalization constant. Actually, comparing our $m/m^{*}(r_s)$ with
that of calculations of Refs.~\onlinecite{Rietschel_Sham} (``RS'') and \onlinecite{Nakano_Ichimaru} (``NI''),
we find that contrary to our results the self-consistent schemes predict a monotonic increase of the inverse
effective mass as a function of $r_s$. Such an increase is important to imply a bandwidth widening at
metallic densities that disagrees with the experimental findings. Thus together with $Z_F$ slightly
``overestimated'' in comparison with the GWA [see inset of Fig.~\ref{invers_eff_mass_Z_2_56}(a)] these
schemes yield the effective mass smaller than that in the noninteracting system, whereas the GWA gives the
reverse. According to Eq.~(\ref{eme_exact}), it means that in the RS and NI cases the momentum derivative of
the self-energy is larger than that evaluated in the GWA. Seemingly, it is caused by a large contribution of
the Hartree-Fock part which is the self-energy for the noninteracting systems providing the effective mass
equal to zero at the Fermi surface. This contribution cannot be canceled by the correlation part of the
self-energy in full measure, as it occurs in the non-self-consistent GWA (see Ref.~\onlinecite{SC_GWA}). As a
result, this leads to an increase of the $m/m^{*}$ ratio. To all appearances, at least for $r_s<4$ we have a
similar situation in the generalized $GW$ self-energy calculations of Ref.~\onlinecite{GV_electron_liquid}.

%=============================================================================================================
\section{\label{sec:conclusions}Conclusions}

In conclusion, we have presented a detailed study of the effect of multiple electron-hole scattering on
quasiparticle properties determined by both the self-energy and its derivatives with respect to momentum and
frequency. To take into account multiple scattering between an electron and a hole in calculations of the
self-energy $\Sigma$, a variational solution of the corresponding Bethe-Salpeter equation obtained in our
preceding paper\cite{IAN_EVC} has been used. This solution representing the $T$ matrix within a local
approximation allows one to sum the self-energy ladder diagrams. To preserve the advantages of the GWA, we
have considered the sum of these diagrams as an additional term (the $T$-matrix contribution $\Sigma^{T}$) to
the $GW$ self-energy $\Sigma^{GW}$. In this approach all weight of the problem is transferred to a form
chosen for the local electron-hole interaction $\widetilde{W}$ that appears in the definition of the $T$
matrix. By examining the irreducible polarizability ladder diagrams, one can identify this interaction with
the exchange part of the many-body local-field factor. Considering this local interaction at the small
four-momentum transfer limit, we have arrived at the expression which gives the results for the exchange
local-field factor in accordance with those known from the literature.

Using the obtained form of $\widetilde{W}$, we have carried out extensive calculations of both the
$\Sigma^{GW}$ and $\Sigma^{T}$ terms and such quasiparticle properties as the damping rate, the quasiparticle
energy, the renormalization constant, and the effective mass enhancement over a broad range of electron
densities in the homogeneous electron gas. The calculations have shown that the $T$-matrix inclusion leads to
an increase of the quasiparticle damping rate especially in the vicinity of $k_F$. This increase depends on
$r_s$ and can exceed the $GW$ prediction by a factor of $1.8$ for $r_s=12$ and $3.6$ for $r_s=56$. Regarding
the question of how a form chosen for the local interaction affects the quasiparticle damping rate, we have
found that the latter is a very form-sensitive quantity, and consequently it is easy to over(under)estimate
the ladder diagrams contribution. We have also revealed that due to small values of the on-shell real part of
$\Sigma^{T}$ the $T$-matrix inclusion modifies slightly the quasiparticle dispersion reducing the $GW$ band
narrowing.

We have found that the $T$-matrix contribution can notably affect the renormalization constant and the
effective mass enhancement. Examining the renormalization constant $Z_F$ as a function of $r_s$ at the Fermi
surface, we have ascertained that in comparison with the $GW$ values the $T$-matrix inclusion reduces $Z_F$.
As a result, the latter becomes closer to the renormalization constant given by the GZ
parametrization\cite{Gori_Ziesche} compatible with QMC calculations of the momentum distribution. A close
analysis of the quasiparticle effective mass $m^{*}/m$ evaluated both in the on-shell approximation and in
the formally exact Dyson scheme (the off-shell approximation) has shown that the on-shell effective mass
depends strongly on $r_s$ and has a divergence\cite{Zhang_et_al} which is shifted by the $T$-matrix inclusion
from $r_s\sim50$ (in the GWA) to $r_s\sim33$. The off-shell $m^{*}/m$ exhibits relatively weak $r_s$
dependence and does not diverge up to the largest $r_s$ considered. In this case, the $T$-matrix contribution
leads only to a slight increase of ``the weight of quasiparticles''.

%=============================================================================================================
\section*{\label{sec:acknowledgments}Acknowledgments}
We would like to thank N.W. Ashcroft and I. Nagy for useful discussions. The work was partially supported by
the Research and Educational Center of Tomsk State University, Departamento de Educaci\'on del Gobierno
Vasco, MCyT (Grant No. FIS 2004-06490-C03-01), and by the European Community 6th framework Network of
Excellence NANOQUANTA (Grant No. NMP4-CT-2004-500198).

%=============================================================================================================
%=============================================================================================================
%=============================================================================================================


\begin{thebibliography}{00}
\bibitem{remark_DOS} At the Fermi surface, the density of states is assumed to be nonzero.
\bibitem{remark_RP} The regular part $\varphi$ is essential only in the integral over remote domain where the
dynamically screened Coulomb interaction is weak.
\bibitem{remark_A_divergence} At that, due to the relation (\ref{small_q_limit}), the local-field
factor and, consiquently, the local interaction $\widetilde{W}$ go towards zero.
\bibitem{remark_SC} We write ``self-consistent'' in quotation marks because it is not implied the full
cycle of recalculations of the self-energy, the Green function, and so on.
\bibitem{remark_Hubbard} This prefactor of ``self-consistency'' is interesting to essentially bring
the Hubbard approximation results closer to those evaluated from Monte Carlo calculations of
Ref.~\onlinecite{CA_PZ}.
\bibitem{remark_magnon} It is necessary to note that the picture may differ drastically in the case of the HEG in the
spin-polarized state where magnon peaks appear in $\mathrm{Im}\Sigma^{T}$.
\bibitem{remark_ferro} Note that the coefficient $A$ used does not meet the condition $\frac{4\alpha r_s}{\pi}A=1$ of a
ferromagnetic instability.
\end{thebibliography}
\end{document}